\documentclass[aps,prb,twocolumn,showpacs,superscriptaddress]{revtex4}

\usepackage{graphicx} 
\usepackage{multirow}
\begin{document}

\title
{Light Transition Metal Monatomic Chains }
\author{C. Ataca}

\affiliation{Department of Physics, Bilkent University, Ankara,
06800, Turkey}
\author{S. Cahangirov}
\affiliation{UNAM-Material Science and Nanotechnology Institute,
Bilkent University, Ankara 06800, Turkey}
\author{ E. Durgun}
\affiliation{Department of Physics, Bilkent University, Ankara,
06800, Turkey}
 \affiliation{UNAM-Material Science and
Nanotechnology Institute, Bilkent University, Ankara 06800,
Turkey}

\author{Y.-R. Jang}\affiliation{Department of Physics, Bilkent University, Ankara,
06800, Turkey}\affiliation{Department of Physics, University of
Incheon, Incheon 402-749, Korea}
\author{ S. Ciraci}\email{ciraci@fen.bilkent.edu.tr}
 \affiliation{Department of Physics, Bilkent
University, Ankara, 06800, Turkey} \affiliation{UNAM-Material
Science and Nanotechnology Institute, Bilkent University, Ankara
06800, Turkey}

\date{\today}

\begin{abstract}
In this paper we investigated structural, electronic and magnetic
properties of \emph{3d} (light) transition metal (TM) atomic
chains using first-principles pseudopotential plane wave
calculations. Periodic linear, dimerized linear and planar zigzag
chain structures and their short segments consisting of finite
number of atoms have been considered. Like Cu, the periodic,
linear chains of Mn, Co and Ni correspond to a local shallow
minimum. However, for most of the infinite periodic chains,
neither linear nor dimerized linear structures are favored; to
lower their energy the chains undergo a structural transformation
to form planar zigzag and dimerized zigzag geometry. Dimerization
in both infinite and finite chains are much stronger than the
usual Peierls distortion and appear to depend on the number of
\emph{3d}-electrons. As a result of dimerization, a significant
energy lowering occurs which, in turn, influences the stability
and physical properties. Metallic linear chain of Vanadium becomes
half-metallic upon dimerization. Infinite linear chain of Scandium
also becomes half-metallic upon transformation to zigzag
structure. An interplay between the magnetic ground state and
atomic as well as electronic structure of the chain has been
revealed. The end effects influence the geometry, energetics and
magnetic ground state of the finite chains. Structure optimization
performed using noncollinear approximation indicates significant
differences from the collinear approximation. Variation of the
cohesive energy of infinite and finite-size chains with respect to
the number of \emph{3d}-electrons are found to mimic the bulk
behavior pointed out by Friedel. The spin-orbit coupling of finite
chains are found to be negligibly small.

\end{abstract}


\maketitle

\section{introduction}

Fabrication of nanoscale structures, such as quantum dots,
nanowires, atomic chains and functionalized molecules have made a
great impact in various fields of science and
technology.\cite{Mokrousov,Tsukagoshi,ciraci1,ciraci2} Size and
dimensionality have been shown to strongly affect the physical and
chemical properties of matter.\cite{Schneider} Electrons in lower
dimensionality undergo a quantization which is different from that
in the bulk materials.\cite{Agrait, Gambardella,Wees} In nanoscale
size, the quantum effects, in particular the discrete nature of
electronic energies with significant level spacing are pronounced.

Suspended monoatomic chains being an ultimate one-dimensional (1D)
nanowire have been produced and their fundemental properties both
theoretically and experimentally investigated.\cite{Sorensen,
Hakkinen, Sanchez, Torres, Okamoto, Yanson, Ohnishi,
Rodrigues,sen1,sen2,tongay} Ballistic electron
transport\cite{Wees} with quantized conductance at room
temperature has been observed in metallic
nanowires.\cite{Okamoto,Yanson} Moreover, magnetic and transport
properties become strongly dependent on the details of atomic
configuration. Depending on the type and position of a foreign
atom or molecule that is adsorbed to a nansostructure, dramatic
changes can occur in the physical properties.\cite{ciraci2} Some
experimental studies, however, aimed at producing atomic chains on
a substrate.\cite{gurlu} Here the substrate-chain interaction can
enter as a new degrees of freedom to influence the physical
properties.

Unlike the metal and semiconductor chains,  not many theoretical
studies are performed on transition metal\cite{Mokrousov2,
Dorantes, Delin3, Bala} (TM) monatomic chains. TM monatomic chains
have the ability to magnetize much more than the bulk.\cite{Delin}
Large exchange interactions of TM atoms in the bulk are overcame
by the large electron kinetic energies, which result in a
nonmagnetic ground state with large bandwidth. On the other hand,
geometries which are nonmagnetic in bulk may have magnetic ground
states in monatomic chains.\cite{Delin} In addition, it is
predicted that the quantum confinement of electrons in metallic
chains should result in a magnetic ground state and also a super
paramagnetic state for some of the TM chains\cite{Wierzbowska} at
finite temperatures. The central issue here is the stability of
the chain and the interplay between 1D geometry and magnetic
ground state.\cite{Mokrousov2, Dorantes}

From the technological point of view, TM monatomic chains are
important in spin dependent electronics, namely
spintronics.\cite{Wolf} While most of the conventional electronics
is based on the transport of information through charges, future
generation spintronic devices will take the advantage of the
electron spin to double the capacity of electronics. It has been
revealed that TM atomic chains either suspended or adsorbed on a
1D substrate, such as carbon nanotubes or Si nanowires can exhibit
high spin-polarity or half-metallic behavior relevant the for
spin-valve effect.\cite{ciraci2} Recently, first-principles
pseudopotential calculations have predicted that the finite size
segments of linear carbon chains capped by specific \emph{3d}-TM
atoms display an interesting even-odd disparity depending on the
number of carbon atoms. For example, CoC$_n$Co linear chain has an
antiferromagnetic ground state for even $n$, but the ground state
changes to ferromagnetic for odd $n$. Even more interesting is the
ferromagnetic excited state of an antiferromagnetic ground state
can operate as a spin-valve when CoC$_n$Co chain is connected to
metallic electrodes from both ends.\cite{durgunsp}

As the chain length decreases, finite-size effects dominate the
magnetic and electronic properties.\cite{Dorantes, Kodama2} When
compared with the infinite case, these are less stable to thermal
fluctuations.\cite{Kachkachi} Additional effects on the behavior
of nanoparticle are their intrinsic properties and the interaction
between them.\cite{Kachkachi, Moruzzi, Kodama2, Luban} Effects of
noncollinear magnetism have to be taken into account as
well.\cite{Gambardella2, Heine, Knickelbein} The end atoms also
exhibit different behavior with respect to the atoms close to the
middle of the structure.\cite{Wallis}

In this paper, we consider infinite, periodic chains of
\emph{3d}-TM atoms having linear and planar zigzag structures and
their short segments consisting of finite number of atoms. For the
sake of comparison, Cu and Zn chains also included in our study.
All the chain structures discussed in this paper do not correspond
to a global minimum, but may belong to a local minimum. The
infinite and periodic geometry is of academic interest and can
also be representative for very long monatomic chains. The main
interest is, however, the short segments comprising finite number
of TM atoms. We examined the variation of energy as a function of
the lattice constant in different magnetic states, and determined
the stable infinite and also finite-size chain structures. We
investigated the electronic and magnetic properties of these
structures. Present study revealed a number of properties of
fundamental and technological interest: The linear geometry of the
infinite, periodic chain is not stable for most of the
\emph{3d}-TM atoms. Even in the linear geometry, atoms are
dimerized to lower the energy of the chain. We found that infinite
linear Vanadium chains are metallic, but become half-metallic upon
dimerization. The planar zigzag chains are more energetic and
correspond to a local minimum. For specific TM chains, the energy
can further be lowered through dimer formation within planar
zigzag geometry. Dramatic changes in electronic properties occur
as a result of dimerization. Magnetic properties of short
monatomic chains have been investigated using both collinear and
noncollinear approximation, which are resulted in different net
magnetic moment for specific chains. Spin-orbit coupling which are
calculated for different direction of applied magnetic field have
been found to be negligibly small.

\section{methodology}

\begin{center}
\begin{figure}
\includegraphics[scale=0.47]{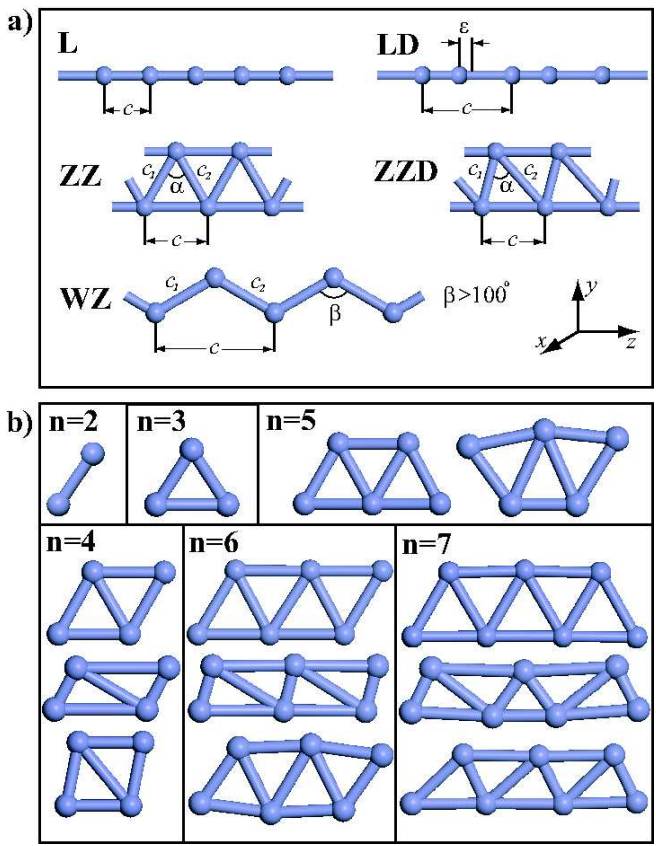}
\caption{(Color online) Various  structures of \emph{3d}-TM atomic
chains. (a) Infinite and periodic structures; L: The infinite
linear monatomic chain of TM atom with lattice constant,
$\emph{c}$. LD: The dimerized linear monatomic chain with two TM
atoms in the cell. $\epsilon$ is the displacement of the second
atom from the middle of the unit cell. ZZ: The planar zigzag
monatomic chain with lattice parameter $\emph{c}$ and unit cell
having two TM atoms. $\emph{c}_{1} \sim \emph{c}_{2}$ and $59^{0}
< \alpha < 62^{0}$. ZZD: The dimerized zigzag structure
$\emph{c}_1 \ne \emph{c}_2$. WZ: The wide angle zigzag structure
$\emph{c}_1 \sim \emph{c}_2$, but $\alpha
> 100^{0}$. (b) Various chain structures of small segments consisting of finite
number $(n)$ of TM atoms, denoted by (TM)$_n$.}\label{fig:1}
\end{figure}
\end{center}

We have performed first-principles plane wave
calculations\cite{payne,vasp} within Density-Functional Theory
(DFT)\cite{kohn} using ultra-soft pseudopotentials.\cite{vander}
We also used PAW\cite{blochl} potentials for noncollinear and
noncollinear spin-orbit calculations of finite chains. The
exchange-correlation potential has been approximated by
generalized gradient approximation (GGA).\cite{gga} For partial
occupancies, we have used the Methfessel-Paxton smearing
method.\cite{methfessel} The width of smearing for infinite
structures has been chosen as 0.1 eV for geometry relaxations and
as 0.01 eV for accurate energy band and density of state (DOS)
calculations. As for finite structures, the width of smearing is
taken as 0.01 eV. We treated the chain structures by supercell
geometry (with lattice parameters $a_{sc}$, $b_{sc}$, and
$c_{sc}$) using the periodic boundary conditions.  A large spacing
($\sim 10 \AA$) between adjacent chains has been assured to
prevent interactions between them. In single cell calculations of
infinite systems, $c_{sc}$ has been taken to be equal to the
lattice constant of the chain. The number of plane waves used in
expanding Bloch functions and \textbf{k}-points used in sampling
the Brillouin zone (BZ) have been determined by a series of
convergence tests. Accordingly, in the self-consistent potential
and total energy calculations the BZ has been sampled by (1x1x41)
mesh points in \textbf{k}-space within Monkhorst-Pack scheme.
\cite{monk} A plane-wave basis set with kinetic energy cutoff
$\hbar^2 |\textbf{k}+\textbf{G}|^2/2m = 350~ eV$ has been used. In
calculations involving PAW potentials, kinetic energy cutoff is
taken as $400~eV$. All atomic positions and lattice constants
($c_{sc}$) have been optimized by using the conjugate gradient
method where total energy and atomic forces are minimized. The
convergence is achieved when the difference of the total energies
of last two consecutive steps is less than $10^{-5}$ eV and the
maximum force allowed on each atom is 0.05 eV/$\AA$. As for finite
structures, supercell has been constructed in order to assure
$\sim 10 \AA$ distance between the atoms of adjacent finite chain
in all directions and BZ is sampled only at the $\Gamma$-point.
The other parameters of the calculations are kept the same. The
total energy of optimized structure ($E_T$) relative to free atom
energies is negative, if they are in a binding state. As a rule,
the structure becomes more energetic (or stable) as its total
energy is lowered. Figure~\ref{fig:1} describes various chain
structures of TM atoms treated in this study. These are infinite
periodic chains and segments of a small number of atoms forming a
string or a planar zigzag geometry. The stability of
structure-optimized finite chains are further tested by displacing
atoms from their equilibrium positions in the plane and
subsequently reoptimizing the structure. Finite-size clusters of
TM atoms are kept beyond the scope of this paper.

\section{Infinite  And Periodic Chain Structures}

\begin{center}
\begin{figure}
\includegraphics[scale=0.45]{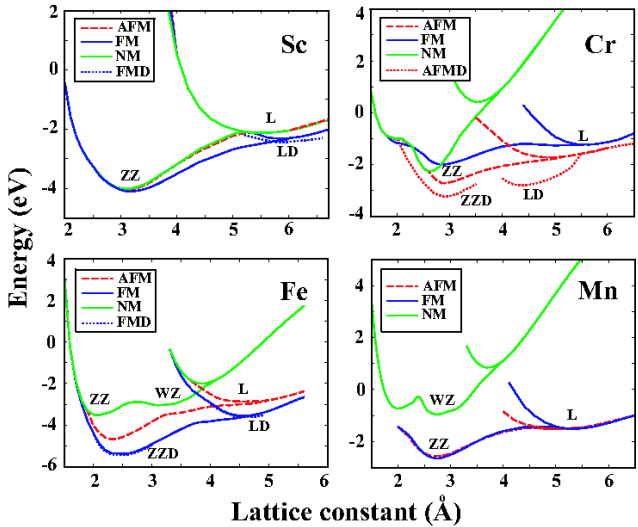}
\caption{(Color online) The energy versus lattice constant, $c$,
of various chain structures in different magnetic states. FM:
ferromagnetic; AFM: antiferromagnetic; NM: nonmagnetic; FMD:
ferromagnetic state in the linear or zigzag dimerized structure;
AFMD: antiferromagnetic state in the dimerized linear or zigzag
structure. The energy is taken as the energy per unit cell
relative to the constituent free atom energies in their ground
state (See text for definition). In order to compare the energy of
the L structure with that of the LD, the unit cell (and also
lattice constant) of the former is doubled in the plot. Types of
structures identified as L, LD, ZZ, ZZD, WZ are describes in
Fig.~\ref{fig:1}. }\label{fig:2}
\end{figure}
\end{center}

\begin{center}
\begin{figure}
\includegraphics[scale=0.45]{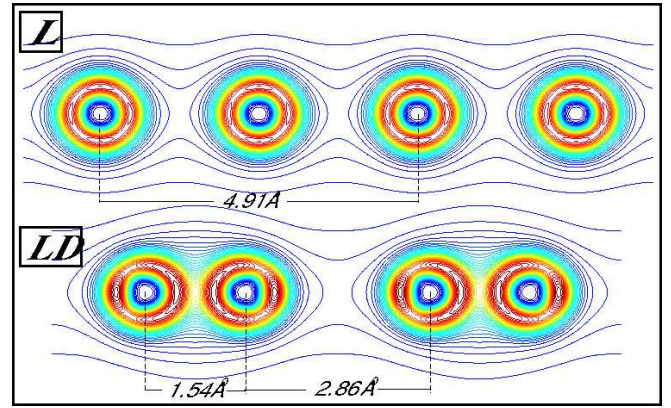}
\caption{(Color online) Charge density contour plots for the
linear (L) and the dimerized linear structure (LD) of Cr monatomic
chains. Interatomic distances are indicated. Contour spacings are
equal to $\Delta \rho=0.0827 e/\AA^{3}$  . The outermost contour
corresponds to $\Delta \rho=0.0827 e/\AA^{3}$ .}\label{fig:8}
\end{figure}
\end{center}

In Fig.~\ref{fig:2}, we present the energy versus lattice constant
of various infinite and periodic chain structures (described in
Fig.~\ref{fig:1}) in different magnetic states. In calculating the
ferromagnetic (FM) state, the structure is optimized each time
using a spin-polarized GGA calculations starting with a different
preset magnetic moment in agreement with Hund's rule. The relaxed
magnetic moment yielding to the lowest total energy has been taken
as the FM state of the chain. For the antiferromagnetic (AFM)
state, we assigned different initial spins of opposite directions
to adjacent  atoms and relaxed the structure. We performed
spin-unpolarized GGA calculations for the nonmagnetic (NM) state.
The energy per unit cell relative to the constituent free atoms is
calculated from the expression, $E=[N E_{a}-E_{T}]$, in terms of
the total energy per unit cell of the given chain structure for a
given magnetic state ($E_{T}$) and the ground state energy of the
free constituent TM atom, $E_{a}$. $N$ is the number of TM atom in
the unit cell, that is $N$=1 for L, but $N$=2 for LD, ZZ, and ZZD
structures. The minimum of $E$ is the binding energy. By
convention $E_{b} < 0$ corresponds to a binding structure, but not
necessary to a stable structure. The cohesive energy per atom is
$E_{c} = -E_{b}/N $. Light transition metal atoms can have
different structural and magnetic states depending on the number
of their \emph{3d} electrons. For example, Sc having a single
\emph{3d} electron, has a shallow minimum corresponding to a
dimerized linear chain structure in the FM state. If the L
structure is dimerized to make a LD structure, the energy of the
chain is slightly lowered. Other linear structures, such as linear
NM, and AFM have higher energy. More stable structure ZZ is,
however, in the FM state. The situation is rather different for
Cr, Fe, and Mn. For example, Cr have LD and more energetic ZZD
structures in the AFM state. It should be noted that in the
dimerized linear chain structure of Cr the displacement of the
second atom from the middle of the unit cell, $\epsilon$, is
rather large. Apparently, the dimerization is stronger than the
usual Peierls distortion. As a result, the nearest neighbor
distance, $(c-\epsilon)$, is much smaller than the second nearest
neighbor distance, $(c+\epsilon)$. This situation poses the
question whether the interaction between the adjacent dimers are
strong enough to maintain the coherence of the chain structure. We
address this question by comparing the energies of individual
dimers with the chain structure. The formation of the LD structure
is energetically more favorable with respect to individual dimer
by $0.54$ eV per atom. Furthermore, the charge density contour
plots presented in Fig.~\ref{fig:8} indicate a significant bonding
between adjacent dimers. Nevertheless, the LD structure has to
transform to more energetic ZZD structure. The zigzag structures
in the AFM, FM and NM states have minima at higher binding
energies and hence are unstable. The linear and LD Fe chains have
a local minimum in FM state. More stable ZZ and ZZD structures in
FM state have almost identical minima in lower binding energy. The
most stable chain structure of Mn among ones described in
Fig.~\ref{fig:2} is ZZ in FM state. It is also saliency to note
that both Fe and Mn chains in NM state undergo a structural
transformation from ZZ to WZ structure. It is noted from
Fig.~\ref{fig:2} that the structure of \emph{3d}-TM atomic chains
are strongly dependent on their magnetic state. Optimized
structural parameters, cohesive energy, magnetic state and net
magnetic moment of infinite linear and zigzag structures are
presented in Table~\ref{tab:1} and Table~\ref{tab:2},
respectively.

\begin{table}

\begin{center}
\begin{tabular}{c||c|c|c|c|c|c|c|c|c|c|}
 & Sc & Ti & V & Cr & Mn & Fe & Co & Ni & Cu & Zn \\
\hline\hline
 $c$ & 6.0 & 4.9 & 4.5 & 4.4 & 2.6 & 4.6 & 2.1 & 2.2 &
2.3 & 2.6 \\ \hline $\varepsilon$ & 0.38 & 0.52 & 0.51 & 0.66 &
0.0 & 0.21 & 0.0 & 0.0 & 0.0 & 0.0 \\ \hline $E_{c}$ & 1.20 & 1.83
&1.86 & 1.40 & 0.76 & 1.81 & 2.10 & 1.99 & 1.54 & 0.15 \\
\hline
MGS & FM & FM & FM & AFM & AFM & FM & FM & FM & NM & NM \\
\hline
 $\mu$ & 1.74 & 0.45 & 1.00 & $\pm$1.95 & $\pm$4.40 & 3.32 & 2.18
& 1.14 & 0.0 & 0.0 \\ \hline
\end{tabular}
\caption{The calculated values for linear structures (L and LD).
The lattice constant, $c$ (in \AA); the displacement of the second
atom in the unit cell of dimerized linear structure, $\epsilon$
(in \AA); cohesive energy, $E_{c}$ (in eV/atom); the magnetic
ground state, MGS; and the total magnetic moment, $\mu$ per unit
cell (in Bohr magnetons, $\mu_{B}$) obtained within collinear
approximation.} \label{tab:1}
\end{center}
\end{table}

\begin{table}
\begin{center}
\begin{tabular}{c||c|c|c|c|c|c|c|c|c|c|}
 & Sc & Ti & V & Cr & Mn & Fe & Co & Ni & Cu & Zn \\
\hline\hline $c$ & 3.17 & 2.60 & 2.60 & 2.90 & 2.76 & 2.40 & 2.30
& 2.30 & 2.40 & 2.50 \\ \hline $c_{1}$ & 2.94 & 2.43 & 1.84 & 1.57
& 2.64 & 2.24 & 2.23 & 2.33 & 2.39 & 2.67 \\ \hline $c_{2}$ & 2.94
& 2.45 & 2.42 & 2.65 & 2.64 & 2.42 & 2.39 & 2.33 & 2.39 & 2.67 \\
\hline $\alpha$ & 65.2 & 64.5 & 73.8 & 82.6 & 63.0& 61.9& 59.6&
59.1 & 60.2 & 55.8 \\ \hline $E_{c}$ &2.05 & 2.78 & 2.64 & 1.57 &
1.32 & 2.69 & 2.91 & 2.74 & 2.16 & 0.37 \\ \hline MGS & FM & FM &
NM & AFM & FM & FM &FM & FM & NM &NM \\ \hline $\mu$ & 0.99 & 0.18
& 0.0 & $\pm$1.82 & 4.36 & 3.19 & 2.05 & 0.92 & 0.0 &0.0\\ \hline
\end{tabular}
\caption{The calculated values for the planar zigzag structures
(ZZ and ZZD): The lattice constant, $c$ (in \AA); the first
nearest neighbor, $c_{1}$ (in \AA); the second nearest neighbor,
$c_{2}$ (in \AA); angle between them, $\alpha$ (in degrees); the
cohesive energy, $E_{c}$ (in eV/atom); the magnetic ground state,
MGS; and the total magnetic moment, $\mu$ per unit cell (in Bohr
magnetons, $\mu_{B}$) obtained within collinear approximation.}
\label{tab:2}
\end{center}
\end{table}

\begin{center}
\begin{figure}
\includegraphics[scale=0.45]{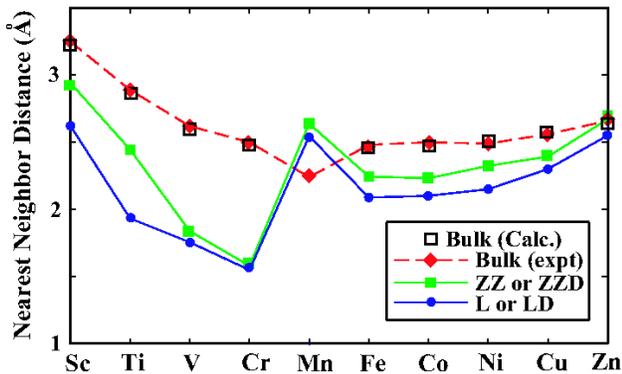}
\caption{(Color online) Variation of the nearest neighbor distance
of \emph{3d}-TM atomic chains and the bulk structures. For the
linear and zigzag structures the lowest energy configuration (i.e.
symmetric or dimerized one) has been taken into account.
Experimental values of the bulk nearest neighbor distances have
been taken from Ref.[\onlinecite{kittel}].}\label{fig:3}
\end{figure}
\end{center}

\begin{center}
\begin{figure}
\includegraphics[scale=0.45]{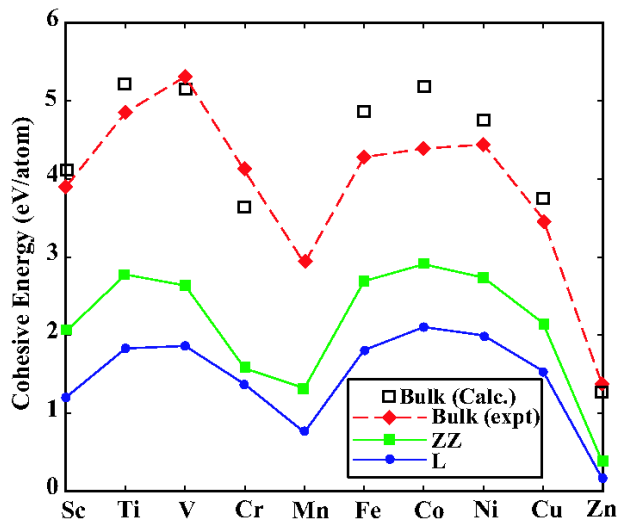}
\caption{(Color online) Variation of the cohesive energy, $E_{c} $
(per atom), of \emph{3d}-TM monatomic chains in their lowest
energy linear, zigzag and bulk structures. For the linear and
zigzag structures the highest cohesive energy configuration (i.e.
symmetric or dimerized one) has been taken into account.
Experimental values of the bulk cohesive energies have been taken
from Ref.[\onlinecite{kittel}]}\label{fig:4}
\end{figure}
\end{center}

 In Fig.~\ref{fig:3} and Fig.~\ref{fig:4} we compare the nearest neighbor distance and the average cohesive energy of the linear and zigzag chain structures with
 those of the bulk metals and plot their variations with respect to their number of \emph{3d} electrons of the TM atom.
 The nearest neighbor distance in the linear and
zigzag structures are smaller than that of the corresponding bulk
structure, but display the similar trend. Namely, it is large for
Sc having a single \emph{3d} electron and decreases as the number
of \emph{3d} electrons, \emph{i.e.} $N_{d}$, increases to four. Mn
is an exception, since the bulk and the chain structure show
opposite behavior. While the nearest neighbor distance of bulk Mn
is a minimum, it attains a maximum value in the chain structure.
Owing to their smaller coordination number, chain structures have
smaller cohesive energy as compared to the bulk crystals as shown
in Fig. ~\ref{fig:4}. However, both L (or LD if it has a lower
energy) and ZZ (or ZZD if it has a lower energy) also show the
well-known double hump behavior which is characteristics of the
bulk TM crystals. Earlier, this behavior was explained for the
bulk TM crystals.\cite{Mokrousov,bethe,friedel} The cohesive
energy of zigzag structures are generally $\sim$ 0.7 eV larger
than that of the linear structures. However, they are 1-2 eV
smaller than that of the bulk crystal. This implies that stable
chain structures treated in this study correspond only to local
minima in the Born-Oppenheimer surface.

\begin{center}
\begin{figure}
\includegraphics[scale=0.65]{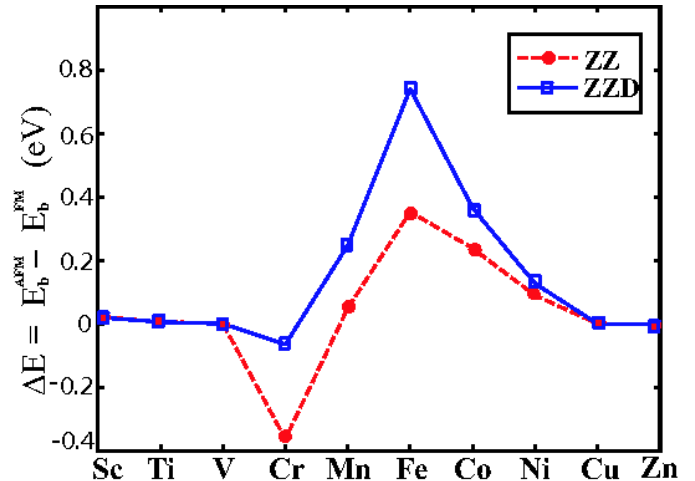}
\caption{(Color online) Variation of the binding energy
difference, $\Delta E$ (per atom) between the lowest
antiferromagnetic and ferromagnetic states of \emph{3d}-TM
monatomic chains. Open squares and filled circles are for the
symmetric zigzag ZZ and dimerized zigzag ZZD chains
respectively.}\label{fig:5}
\end{figure}
\end{center}

We note that spin-polarized calculations are carried out under
collinear approximation. It is observed that all chain structures
presented in Table~\ref{tab:1} and Table \ref{tab:2} have magnetic
state if $N_{d} < 9$. Only Cr and Mn linear chain structures and
Cr zigzag chain structure have an AFM lowest energy state. The
binding energy difference between the AFM state and the FM state,
$\Delta E=E_{b}^{AFM}-E_{b}^{FM}$, is calculated for all light
TMs. Variation of $\Delta E$ with $N_{d}$ is plotted in
Fig.~\ref{fig:5}. We see that only Cr ZZ and ZZD chains have an
AFM lowest energy state. $\Delta E$ of Fe is positive and has the
largest value among all \emph{3d}-TM zigzag chains. Note that
$\Delta E$ increases significantly as a result of dimerization.

\begin{center}
\begin{figure}
\includegraphics[scale=0.45]{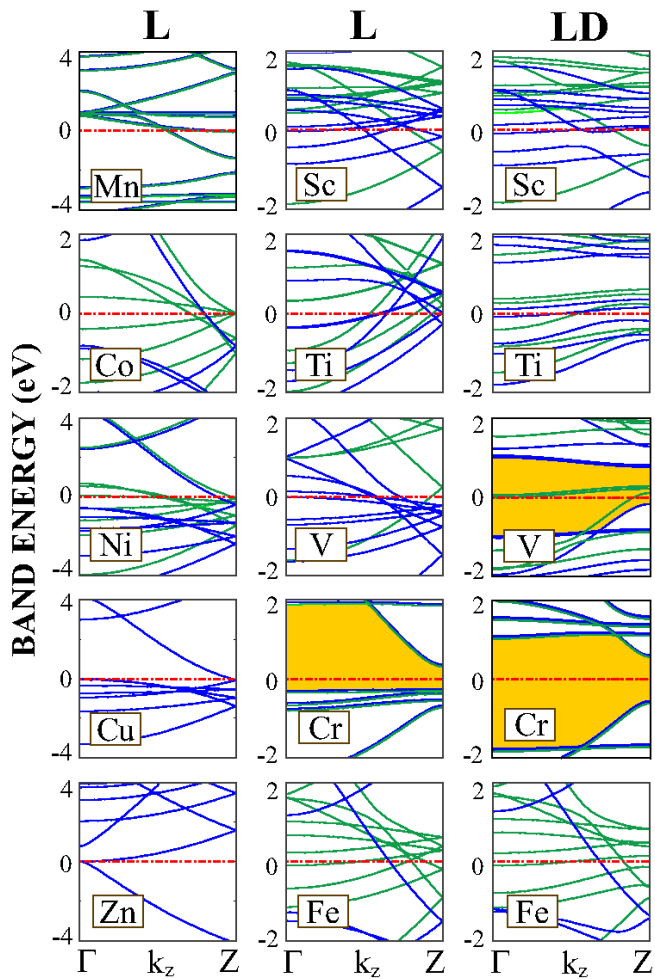}
\caption{(Color online) Energy band structures of \emph{3d}-TM
atomic chains in their L and LD structures. The zero of energy is
set at the Fermi level. Gray and black lines are minority and
majority bands, respectively. In the antiferromagnetic state
majority and minority bands coincide. Energy gaps between the
valence and the conduction bands are shaded.}\label{fig:6}
\end{figure}
\end{center}

\begin{center}
\begin{figure}
\includegraphics[scale=0.45]{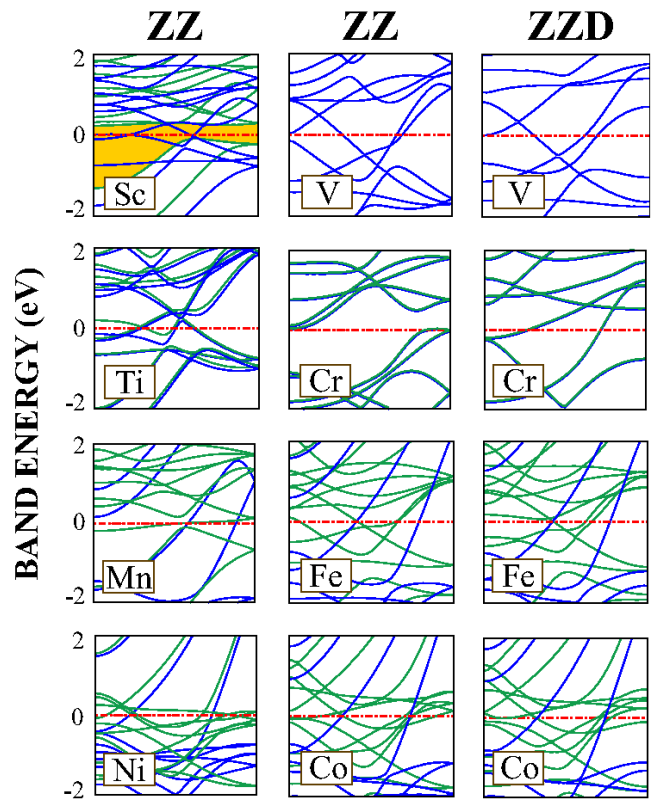}
\caption{(Color online) Energy band structures of \emph{3d}-TM
atomic chains in their zigzag (ZZ) and dimerized zigzag (ZZD)
structures. The zero of energy is set at the Fermi level. Gray and
black lines are minority and majority spin bands, respectively.
The gray and dark lines coincide in the antiferromagnetic state.
Only dark lines describe the bands of nonmagnetic state. The
energy gap between the valence and the conduction bands is
shaded.}\label{fig:7}
\end{figure}
\end{center}

Having discussed the atomic structure of \emph{3d}-TM chains, we
next examine their electronic band structure. In Fig.~\ref{fig:6},
the chain structures in the first column do not dimerize. The
linear chains placed in the third column are dimerized and changed
from the L structure in the second column to form the LD
structure. Most of the linear structures in Fig.~\ref{fig:6}
display a FM metallic character with broken spin degeneracy. A few
exceptions are Mn, Cr, and V chains. The linear Mn chain has an
AFM state, where spin-up (majority) and spin-down (minority) bands
coincide. Chromium L and LD structures are AFM semiconductors.
Vanadium is a ferromagnetic metal for both spins, but becomes
half-metallic upon dimerization. In half-metallic state, the chain
has integer number of net spin in the unit cell. Accordingly,
Vanadium chain in the LD structure is metallic for one spin
direction, but semiconducting for the other spin direction. Hence,
the spin polarization at the Fermi level, \emph{i.e.}
$P=[D_{\uparrow}(E_F)-D_{\downarrow}(E_F)]/[D_{\uparrow}(E_F)+D_{\downarrow}(E_F)]$
is $100\%$. Bands of Cu and Zn with filled \emph{3d}-shell in
nonmagnetic state are in agreement with previous
calculations.\cite{Tung} In Fig.~\ref{fig:7}, the chain structures
in the first column have only ZZ structure. The zigzag chains in
the second column are transformed to a lower energy (\emph{i.e.}
more energetic) ZZD structure in the third column. The ZZ chain of
Sc is stable in a local minimum and displays a half-metallic
character with 100\% spin-polarization at the Fermi level.
Accordingly, a long segment of ZZ chain of Sc can be used as a
spin-valve. Ti, Mn, and Ni in their stable zigzag structures are
FM metals. The stable ZZD structure of Fe and Co chains are also
FM metals. The ZZ and relatively lower energy ZZD structure of V
chain are nonmagnetic. Both ZZ and ZZD structures of Cr are in the
AFM state.

For Co and Fe in the ZZD structure more bands of one type of spin
cross the Fermi level as compared to those of the other type of
spin resulting in a high spin-polarization at the Fermi level.
This situation implies that in the ballistic electron transport,
the conductance of electrons with one type of spin is superior to
electrons with the opposite type of spin; namely $\sigma
_{\downarrow} \gg \sigma _{\uparrow}$. Accordingly, the
conductance of electrons across the Fe and Co chains becomes
strongly dependent on their spin-directions. This behavior of the
infinite periodic Fe or Co chain is expected to be unaltered to
some extend for long, but finite chains and can be utilized as a
spin-dependent electronic device. In closing this section, we
emphasized that the infinite, periodic chains of \emph{3d}-TM
atoms can be in the zigzag structure corresponding to a local
minimum. However, most of the zigzag structures are dimerized.
Dimerization causes remarkable changes in electronic and magnetic
properties.

\section{Short Chain Structures}

Periodic infinite chains in Section III are only ideal structures;
long finite-size segments perhaps can attain their physical
properties revealed above. On the other hand, the end effects can
be crucial for short segments consisting of few atoms which may be
important for various spintronic applications. In this section, we
examine short segments of \emph{3d}-TM chains consisting of $n$
atoms, where $n$=2-7.

\subsection{Collinear Approximation}

We first study the atomic structure and magnetic properties of the
finite chains within collinear approximation using ultra-soft
pseudopotentials.\cite{vander} The linear structure is unstable
for the finite size segments. Various planar zigzag structures,
which are only a local minima, are described in Fig.~\ref{fig:1}.
We optimized the geometry of these zigzag structures with
different initial conditions of magnetic moment on the atoms
within collinear approximation. If the final optimized structures
for $q$ different initial conditions result in different average
cohesive energy (or different total energy), they may actually
trapped in different local minima. Here we considered following
different initial conditions: (1) At the beginning, opposite
magnetic moments, $\pm\mu_a$, have been assigned to adjacent
atoms, and the total magnetic moment, $\mu = \sum  \mu_a$, has
been forced to vanish at the end of optimization for $n=2-7$.
Initial magnetic moment, $\mu_a$, on atoms are determined from the
Hund's rule. (2) For $n=2-7$, initial magnetic moment of all atoms
have been taken in the same direction, but the final magnetic
moment of the structure has been determined after optimization
without any constraint. (3) For $n=2-7$, the system is relaxed
using spin-unpolarized GGA. (4) For $n=2-7$, initial magnetic
moment of chain atoms have been assigned as is done in (1), but
$\mu = \sum \mu_a$ is not forced to vanish in the course of
relaxation. (5) For $n=2-7$, spin-polarized GGA calculations have
been carried out without assigning any initial magnetic moment.
(6) We have assigned the magnetic moment
$\uparrow\downarrow~\downarrow\uparrow $ for $n=4$, and
$\uparrow\downarrow~\downarrow~\downarrow\uparrow $ for $n=5$.
Here different spacings between two spin-arrows indicate different
bond lengths. This way different exchange coupling for different
bond lengths and hence dimerization is accounted. (7) The initial
magnetic moment on atoms
$\uparrow\downarrow~\downarrow\uparrow~\uparrow\downarrow $ for
$n=6$ and
$\uparrow\downarrow\uparrow~\uparrow~\uparrow\downarrow\uparrow $
for $n=7$ have been assigned. (8) Similar to (7), initial magnetic
moment $ \uparrow\downarrow\uparrow~\uparrow\downarrow\uparrow$
and $
\uparrow\downarrow~\downarrow\uparrow\downarrow~\downarrow\uparrow$
have been assigned for n=6 and n=7, respectively. Last three
initial conditions are taken into consideration due to the fact
that different bond lengths of \emph{3d}-TM atoms affect the type
of magnetic coupling between consecutive atoms.\cite{Mokrousov2}
The initial atomic structures have been optimized for these
initial conditions except Cu and Zn. Only first three conditions
are consistent with Cu and Zn. As the initial geometry, a segment
of $n$ atoms has been extracted from the optimized infinite zigzag
chain and placed in a supercell, where the interatomic distance
between adjacent chains was greater than $10$ \AA~ for all atoms.
Our results are summarized in Table~\ref{tab:3}, where the
magnetic orders having the same lowest total energy occurred $p$
times from $q$ different initial conditions, are presented. In
this respect the magnetic ordering in Table~\ref{tab:3} may be a
potential candidate for the magnetic ground state.

\begin{widetext} \centering
\begin{table*}
\caption{The average cohesive energy $E_{c}$ (in eV/atom); the net
magnetic moment $\mu$, (in Bohr magneton $\mu_{B}$); magnetic
ordering (MO); LUMO-HOMO gap of majority/minority states,
$E^{\uparrow}_{G}$ and $E^{\downarrow}_{G}$, respectively (in eV)
for lowest energy zigzag structures. $p/q$ indicates that the same
optimized structure occurred $p$ times starting from $q$ different
initial conditions.(See text) Results have been obtained by
carrying out structure optimization within collinear approximation
using the ultra-soft pseudopotentials.} \label{tab:3}

\begin{tabular}{c|c||c|c|c|c|c|c|c|c|c|c|}
\multicolumn{2}{c||}{ZZ} & Sc & Ti & V & Cr & Mn & Fe & Co & Ni &
Cu & Zn \\ \hline\hline \multirow{5}{*}{$n=2$} & $E_{c}$ & 0.83
&1.38 &1.29 &   0.93 &   0.32 &   1.29 &   1.49 &   1.38& 1.14&
0.02 \\
&$\mu$&4.0  &  2.0  &  2.0&    0.0 &   10.0 &  6.0 &   4.0&
2.0& 0.0 &   0.0 \\
&$E^{\uparrow}_G$/$E^{\downarrow}_G$  & 0.59/1.60 &0.29/1.01
&1.03/1.22 & 2.17/2.17&2.04/0 &1.14/0.59 & 1.42/0.36&
1.48/0.27 &1.59/1.59 &3.96/3.96 \\
&MO&  FM & FM &  FM &  AFM& FM & FM & FM&  FM&
NM &NM \\
&$(p/q=5)$&2 &   1  &  2 &   1 &   1 & 2& 2 &1 & 3 & 3
\\ \hline

\multirow{5}{*}{$n=3$}&$E_{c}$ &1.30  &  1.87 &   1.61  &  0.91&
0.63 &1.72&    1.84 &   1.78 &   1.24  &  0.12 \\
&$\mu$&1.0  &  6.0  &  3.0  &  6.0  &  15.0 &  10.0 &   7.0&
2.0& 1.0&    0.0 \\
&$E^{\uparrow}_G$/$E^{\downarrow}_G$&0.66/0.44 &  0.45/1.08  &
0.31/0.78 & 1.23/2.03 & 1.66/0.35 & 0.39/0.58&
0.19/0.18 &  0.87/0.24  & 0.08/1.55 &  2.96/2.96 \\
&MO&FM & FM  &FM & FM &  FM & FM & FM & FM &
FM& NM \\
&$(p/q=5)$&3  &  1  &  2 &   2  &  1  &  2& 3& 3 & 1 & 3
\\ \hline

\multirow{5}{*}{$n=4$}&$E_{c}$ &1.54   & 2.13   & 2.01  &
1.16&0.84& 2.07  &  2.31   & 2.08 &1.61 &   0.13 \\
&$\mu$&4.0 &   2.0   & 2.0  & 0.0   & 18.0  & 14.0 &
10.0&4.0& 0.0 &   0.0 \\
&$E^{\uparrow}_G$/$E^{\downarrow}_G$&0.37/0.36 &  0.46/0.50 &
0.35/0.30 & 1.16/0.61&1.16/0.50& 1.47/0.04 &  1.98/0.34 &  1.10/0.25 &  0.96/0.96 &  2.35/2.35 \\
&MO&FM & FM & FM & AFM* & FM & FM & FM & FM & NM & NM \\
&$(p/q=6)$&5 &   3 &   4 &   4  &  4 &   2& 1 &3 & 3 & 3
\\ \hline

\multirow{5}{*}{$n=5$}&$E_{c}$&1.63   & 2.27  &  2.08  &
0.83&0.91& 2.25 &   2.46  &  2.23  &  1.74 &   0.15 \\
&$\mu$&3.0  &  0.0  &  0.0 &   0.0   & 5.0   & 16.0 &
11.0&6.0& 1.0  &  0.0 \\
&$E^{\uparrow}_G$/$E^{\downarrow}_G$&0.29/0.46&   0.43/0.43 &
0.49/0.40 & 0.47/0.52&1.12/0.30& 1.42/0.56 &  1.53/0.37&   1.47/0.09& 1.42/0.90&   1.96/1.96 \\
&MO&FM&  AFM* &AFM* &AFM* & FM &  FM & FM & FM & FM &
NM \\
&$(p/q=6)$& 3  &  4  &  2  &  4 &   4 &   1& 1 &3 & 1 & 3
\\ \hline

\multirow{5}{*}{$n=6$}&$E_{c}$& 1.69  &  2.32  &  2.26 &
1.29&1.02 &2.31 & 2.50   &   2.29  &  1.75   & 0.17 \\
&$\mu$&8.0  &  0.0  &  0.0 &   0.0 &   0.0  &
20.0&14.0 &6.0 &2.0   & 0.0 \\
&$E^{\uparrow}_G$/$E^{\downarrow}_G$& 0.22/0.29  & 0.44/0.44&
0.54/0.54 & 0.53/0.55& 0.41/0.38& 1.33/0.41&0.30/0.32& 0.28/0.10& 1.42/0.95 &  1.88/1.88 \\
&MO& FM & AFM &AFM &AFM*  &AFM* & FM &FM &  FM & FM & NM \\
&$(p/q=7)$& 3   & 3  &  7 &   4 &   4  &  2& 4& 4 & 1 & 3
\\ \hline

\multirow{5}{*}{$n=7$}&$E_{c}$&1.74 &   2.38 &   2.22 &   1.25&
1.06& 2.35 &   2.58  &  2.36    &1.84 &   0.18 \\
&$\mu$&7.0  &  6.0 &   5.0 &   6.0  &  5.0 &  22.0 & 15.0&
8.0& 1.0 &   0.0 \\
&$E^{\uparrow}_G$/$E^{\downarrow}_G$&0.01/0.33 &  0.34/0.21 &
0.32/0.48 & 0.54/0.68&0.85/0.42& 0.95/0.29 &  0.98/0.17 &  0.83/0.09  & 0.79/0.61  & 1.77/1.77 \\
&MO& FM  &FM & FM & FM & FM & FM & FM & FM & FM&
NM \\
&$(p/q=7)$&5  & 3 &  6  & 4 &  5 &  1 &  2 &  4 &  1  & 3 \\
\hline

\end{tabular}

\end{table*}
\end{widetext}

\begin{center}
\begin{figure}
\includegraphics[scale=0.55]{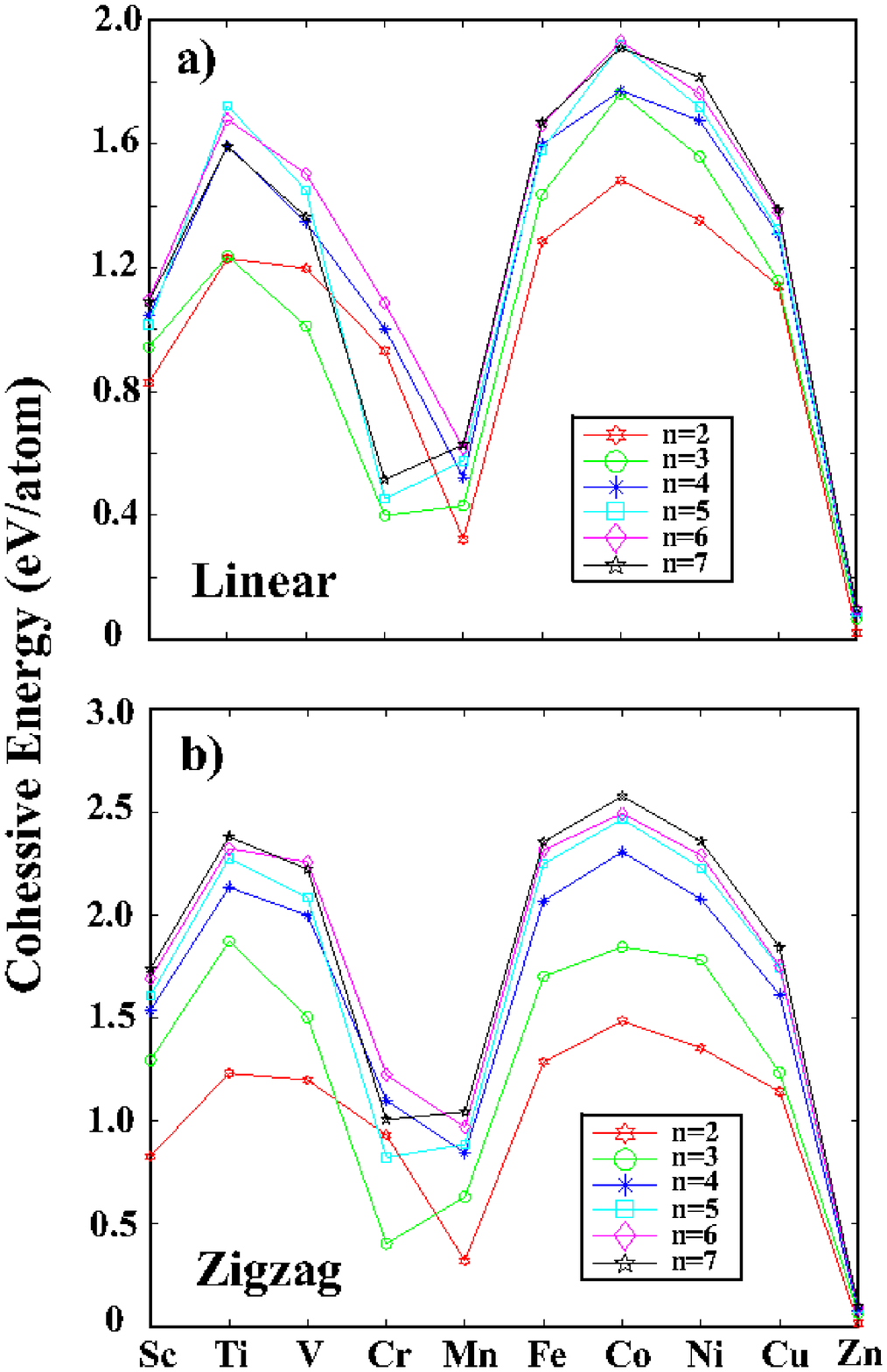}
\caption{(Color online) Variation of the average cohesive energy
of small segments of chains consisting of $n$ atoms. (a) The
linear chains; (b) the zigzag chains . In the plot, the lowest
energy configurations for each case obtained by optimization from
different initial conditions.}\label{fig:9}
\end{figure}
\end{center}

The average cohesive energy of finite-size chains increase with
increasing $n$. In Fig.~\ref{fig:9}, we plot the average cohesive
energy of these small segments consisting of $n$ atoms. For the
sake of comparison, we presented the plots for the linear and
zigzag structures. The average values of cohesive energy in
Fig.~\ref{fig:9}(b) are taken from Table~\ref{tab:3}. We note
three important conclusions drawn from these plots. (i) The
cohesive energies of the zigzag structures are consistently larger
than those of the linear structures, and the cohesive energies
also increase with increasing $n$. (ii) For each types of
structures, as well as for each $n$, the variation of $E_{c}$ with
respect to the number of \emph{3d} electrons in the outer shell,
$N_{d}$, exhibits a double hump shape, which is typical of the
bulk and the infinite chain structures as presented in
Fig.~\ref{fig:4}. (iii) For specific cases $E_c(n_2)<E_c(n_1)$,
even if $n_2>n_1$ (V and Cr). This situation occurs because energy
cannot be lowered in the absence of dimerization.

Most of the finite zigzag chain structure of \emph{3d}-TM atoms
have a FM lowest energy state. The magnetic ordering specified by
AFM* for specific chains indicates that the magnetic moment on
individual atoms, $\mu_a$, may be in opposite directions or may
have unequal magnitudes, but the total magnetic moment, $\mu =
\sum \mu_a$, adds up to zero.  The finite chains of Zn atom are
always nonmagnetic for all $n$. Finite zigzag chains of Cu are
nonmagnetic for even $n$, except $n=6$. Interestingly, the
dimerized linear chain of Cr ($n=5$) with a FM lowest energy state
is more energetic than that of the zigzag chain given in
Table~\ref{tab:3}. The geometry of this structure is such that two
dimers consisting of two atoms are formed at both ends of the
linear structure and a single atom at the middle is located
equidistant from both dimers. The distance from the middle atom to
any of the dimers is approximately twice of the distance between
the atoms in the dimer. Even though the nearest neighbor distance
of the middle atom to dimers is long, there is a bonding between
them. The cohesive energy is $\sim0.2 $ eV higher than that of the
zigzag case, and the total magnetic moment of the structure
($6~\mu_{B}$) is provided by the atom at the middle. This is due
to the fact that two dimers at both ends are coupled in the AFM
order. This is an expected result because the cohesive energy (per
atom) of Cr$_2$ is higher than that of Cr$_5$ in the zigzag
structure. The LUMO/HOMO gap for majority and minority spin states
usually decrease with increasing $n$. However, depending of the
type of TM atom, the maximum value of the gap occurs for different
number of atoms $n$. The zigzag chain of Zn atoms usually have the
largest gap for a given $n$.

Even though the total magnetic moment, $\mu=\sum\mu_a$, of the
AFM* state vanishes for the finite molecule, LUMO-HOMO gaps for
majority and minority states are not generally the same as in the
AFM state. This can be explained by examining the magnetic moment
on every individual atom and the geometry of the molecule. For
Cr$_4$, the magnetic moment on each atom are lined up as described
in the sixth initial condition. In this ordering, two dimers each
consisting of two atoms are in the AFM ordering within themselves,
but in the FM ordering with each other. The distribution of final
magnetic moment on atoms for Mn$_6$ also obey one of the initial
conditions (case 7). Three dimers each consisting of two atoms
coupled in the AFM order within themselves, but in the FM order
with each other. Similar results are also obtained for other AFM*
states.

The zigzag planar structure for $n>3$ in Table~\ref{tab:3}
corresponds to a local minimum. To see whether the planar zigzag
structures are stable, or else it transforms to other geometry by
itself is a critical issue. To assure that the finite chain
structures of $n=4$ and $n=7$ in Table~\ref{tab:3} are stable in a
local minimum, we first displaced the atoms out of planes, then we
optimized the structure. Upon relaxation all displaced atoms
returned to their equilibrium position on the plane.

\subsection{Noncollinear approximation and the spin-orbit interaction}

\begin{widetext}
\centering
\begin{table*}
\caption{The average cohesive energy, $E_c$(in eV/atom); the
components $(\mu_x,~\mu_y,~\mu_z)$ and the magnitude of the net
magnetic moment $\mu$ (in $\mu_B$); LUMO-HOMO gap $E_G$ (in eV) /
energy gap under $1~\mu_B$ applied magnetic field along
\emph{x}-direction(/\emph{z}-direction) $E_G^x/E_G^z$; magnetic
ordering MO; Spin-orbit coupling energy $\triangle
E_{so}^x(/\triangle E_{so}^z)$ (in meV) under applied magnetic
field along \emph{x}-direction(/\emph{z}-direction). $p/q$
indicates that the same optimized structure occurred $p$ times
starting from $q$ different initial contidions. Results have been
obtained by carrying out structure optimization calculations
within noncollinear approximation using PAW potentials. Mn$_7$ is
not stable in the planar ZZ structure.} \label{tab:4}

\begin{tabular}{c|c||c|c|c|c|c|c|c|}
\multicolumn{2}{c||}{ZZ} & Sc & Ti & V & Cr & Mn &  Co & Ni \\
\hline\hline \multirow{7}{*}{$n=2$} & $E_{c}$ &0.85 &1.55&
1.57 &0.52 &0.45&      1.49 & 1.57 \\
&$(\mu_x,\mu_y,$& (    2.3,    2.3,& ( 0.0, 0.0, & ( 1.2, 1.2, & (
0.0, 0.0, & ( 5.8, 4.9,  & ( 2.8,
2.8, &(1.7,    1.0, \\
&$\mu_z),~\mu$ &2.3), 4.0&2.0), 2.0&1.2), 2.0 &0.0), 0.0&6.6), 10.0&0.0), 4.0&0.0), 2.0\\
&$E_G/E_G^x/E_G^z$&0.49/0.18/0.17& 0.36/0.36/0.36& 0.67/0.67/0.66&
0.56/1.87/1.87& 0.18/0.18/0.18& 0.05/0.05/0.05& 0.18/0.17/0.3\\
&MO&FM&FM&FM&AFM&FM&FM&FM \\
&$\triangle E_{so}^x/\triangle E_{so}^z$&3.60/3.80&4.70/3.90&8.30/8.00&10.90/10.90&13.10/13.30&0.01/9.90&33.30/32.50\\
&$(p/q=5)$&4&4&5&3&3&5&4\\
\hline

\multirow{7}{*}{$n=3$} & $E_{c}$ &1.36 & 2.00&1.90
 &0.71 &0.68&     1.89  &2.01  \\
&$(\mu_x,\mu_y,$&(0.7, 0.7, &(2.2, 2.2,  &(0.6, 0.6,  &(5.6, 2.2,  & (1.4, -2.6,  & (0.3, 0.7, & (1.2, 1.2,  \\
&$\mu_z),~\mu$ &-0.3), 1.0&2.5), 4.0&0.6), 1.0 &0.0), 6.0&0.0), 3.0&7.0), 7.0&1.1), 2.0\\
&$E_G/E_G^x/E_G^z$&0.37/0.37/0.37&0.26/0.26/0.25&0.44/0.44/0.44&1.01/1.01/1.01&0.25/0.24/0.24& 0.34/0.11/0.12&0.11/0.11/0.10 \\
&MO&FM&FM&FM&FM&FM&FM&FM \\
&$\triangle E_{so}^x/\triangle E_{so}^z$&3.70/3.70&4.70/4.70& 8.40/8.40& 10.40/10.50& 13.10/13.00&8.20/9.60&33.10/32.70\\
&$(p/q=5)$&4&2&3&2&1&1&5\\
\hline

\multirow{7}{*}{$n=4$} & $E_{c}$ &1.60 &2.36 &2.34
 &0.88 &1.01&      2.28 &2.30  \\
&$(\mu_x,\mu_y,$&(0.6, 1.7, &(1.2, 1.2,  &(0.0, 0.0  & (0.0, 0.0 & (0.0, 0.0,  &(4.6, 4.6, & (-0.8, -2.0  \\
&$\mu_z),~\mu$ &0.9), 2.0&1.2), 2.0& 0.0), 0.0&0.0), 0.0&0.0), 0.0&4.6), 8.0&3.4), 4.0\\
&$E_G/E_G^x/E_G^z$&0.29/0.29/0.29&0.41/0.41/0.41&0.28/0.28/0.28&1.09/1.09/1.09&0.30/0.30/0.30& 0.03/0.03/0.03&0.06/0.21/0.20 \\
&MO&FM&FM&AFM&AFM&AFM&FM&FM \\
&$\triangle E_{so}^x/\triangle E_{so}^z$&3.70/3.70&4.70/4.70&8.40/8.40&10.30/10.20&13.20/13.20&8.30/8.80&32.10/32.20\\
&$(p/q=5)$&3&4&2&1&3&5&2\\
\hline

\multirow{7}{*}{$n=5$} & $E_{c}$ &1.67 &2.48 &2.46
 &0.99 &1.21&     2.49  & 2.30 \\
&$(\mu_x,\mu_y,$&(0.8, 0.2 & (0.0, 0.0 &(0.7, 0.5  &(2.5, 2.45,  &(-1.3, 1.8,  & (-2.4, 10.6, &(2.0, 3.3,   \\
&$\mu_z),~\mu$ &0.6), 1.0&0.0), 0.0&0.6), 1.0 &1.9), 4.0&-2.0), 3.0&-1.4), 11.0&1.1), 4.0\\
&$E_G/E_G^x/E_G^z$&0.26/0.26/0.26&0.34/0.34/0.34&0.27/0.27/0.27&0.28/0.44/0.44&0.09/0.21/0.21& 0.33/0.33/0.33&0.21/0.20/0.20 \\
&MO&FM&AFM&FM&FM&FM&FM&FM \\
&$\triangle E_{so}^x/\triangle E_{so}^z$&3.80/3.50&4.80/4.80&8.20/8.20&10.40/10.40&14.10/13.00&8.90/8.90&32.40/31.80\\
&$(p/q=5)$&4&4&3&1&1&2&5\\
\hline

\multirow{7}{*}{$n=6$} & $E_{c}$ &1.74 & 2.53&2.57
 &1.24 &1.30&     2.55  & 2.45 \\
&$(\mu_x,\mu_y,$&(0.0, 0.0, &(0.0, 0.0,  & (0.0, 0.0, & (0.0, 0.0, & (0.0, 0.0,  & (6.73, 6.73,  &(0.0, 0.0,   \\
&$\mu_z),~\mu$ &0.0), 0.0&0.0), 0.0& 0.0), 0.0&0.0), 0.0&0.0), 0.0&7.31), 12.0&0.0), 0.0\\
&$E_G/E_G^x/E_G^z$&0.19/0.19/0.19&0.32/0.32/0.32&0.38/0.38/0.37&0.77/0.77/0.77&0.48/0.48/0.48& 0.20/0.20/0.20&0.20/0.17/0.17 \\
&MO&AFM&AFM&AFM&AFM&AFM&FM&AFM \\
&$\triangle E_{so}^x/\triangle E_{so}^z$&3.70/3.70& 4.70/4.70& 8.10/8.10& 10.30/10.30& 13.20/13.30 &8.00/8.40 & 32.30/32.30\\
&$(p/q=5)$&1&4&5&1&4&5&5\\
\hline

\multirow{7}{*}{$n=7$} & $E_{c}$ &1.81 &2.60 &2.56
 &1.13 &&     2.64  &2.57  \\
&$(\mu_x,\mu_y,$& (5.2, 5.2,& (1.1, 2.8,  & (0.6, 0.6,  & (-0.2, 0.6,   & &(8.7, 8.7,  &(4.7, 4.6   \\
&$\mu_z),~\mu$ &5.2), 9.0& 0.0), 3.0&0.6), 1.0 &6.0), 6.0&&8.7), 15.0&4.6), 8.0\\
&$E_G/E_G^x/E_G^z$&0.15/0.15/0.15&0.19/0.19/0.20&0.23/0.23/0.23&0.39/0.39/0.39&&0.09/0.09/0.09&0.01/0.05/0.05 \\
&MO&FM&FM&FM&FM&&FM&FM \\
&$\triangle E_{so}^x/\triangle E_{so}^z$&3.80/3.80&4.90/4.80&8.20/8.20& 10.60/10.40&&8.30/8.50& 32.50/32.20\\
&$(p/q=5)$&2&1&5&2&&5&5\\
\hline

\end{tabular}
\end{table*}
\end{widetext}

In cases where both AFM and FM couplings occur and compete with
each other, collinear magnetism fails for modelling the ground
state magnetic ordering. A midway between AFM and FM exchange
interactions results in allowing the spin quantization axis to
vary in every site of the structure. Geometric structure also
influences noncollinear magnetism. Frustrated antiferromagnets
having triangular lattice structure, disordered systems, broken
symmetry on the surface will result in noncollinear magnetism.
$\alpha$-Mn, spin glasses, domain walls, Fe clusters are examples
of this type. Coupling the magnetic moment to the crystal
structure (spin-orbit coupling) poses the magnetic anisotropy
which again results in the noncollinear magnetism in the
structure. Finite structures that is studied in this paper, all
have low symmetry and AFM-FM coupling competition which increase
the probability of observing noncollinear magnetism. There are
many different approaches for implementing noncollinear magnetism,
such as ASW (Augmented Spherical Wave), CPA (Coherent-Potential
Approximation), LSDA (Local-Spin-Density Approximation). In our
study, we use fully unconstrained approach to noncollinear
magnetism.\cite{1} The Hamiltonian of the system after making
simplifications and approximations will be:

\begin{eqnarray}\label{Hamiltonian}
H^{\alpha\beta}[n,\{\textbf{R}\}]& = &
-\frac{1}{2}\delta_{\alpha\beta}+\tilde{v}^{\alpha\beta}_\textrm{eff}+
\nonumber \\
& &
\sum_{(i,j)}|\tilde{p_i}>(\hat{D}^{\alpha\beta}_{ij}+^1D^{\alpha\beta}_{ij}
-^1\tilde{D}^{\alpha\beta}_{ij})<\tilde{p_j}| \nonumber
\end{eqnarray}

Here $\tilde{v}^{\alpha\beta}_\textrm{eff}$ is the effective
one-electron potential which depends on the electron density and
the magnetization at each site;
$-$$1\over{2}$$\delta_{\alpha\beta}$ stands for the kinetic energy
of the system. $\hat{D}^{\alpha\beta}_{ij}$,
$^1D^{\alpha\beta}_{ij}$ and $^1\tilde{D}^{\alpha\beta}_{ij}$
terms in the summation sign over augmented channels represent the
correction terms for long range, effective potential and wave
functions. For further details  see Ref.[\onlinecite{1,2,3,4,5,6}]

The finite chains discussed in the previous section within
collinear approximation will now be treated using noncollinear
approximation. To this end, the structure of chains have been
optimized starting from the same initial geometry (starting from a
segment of $n$ atoms extracted from the optimized infinite ZZ (or
ZZD) chain placed in a supercell) and following five different
initial configurations of spins on individual atoms. (i) The
direction of the initial magnetic moment on the atoms are
consecutively altered as $xyzy$. (ii) No preset directions are
assigned to the individual atoms, they are determined in the
course of structure optimization using noncollinear approximation.
(iii) For each triangle, the initial magnetic moment on the atoms
have a non-zero component only in the $xy$-plane, but
$(\sum\limits_{\triangle}\mu_a)_{xy}=0 $. (Here `$\triangle$'
stands for the summation over the atoms forming a triangle) (iv)
Similar to (iii), but $(\sum\limits_{\triangle}\mu_a)_z\neq0$. (v)
In a zigzag chain, the magnetic moment of atoms on the down row
are directed along \emph{z}-axis, while those on the up row are
directed in the opposite direction. Using these five different
initial conditions on the magnetic moment of individual atoms, the
initial atomic structure is optimized using both
ultra-soft\cite{vander} pseudopotentials and PAW\cite{blochl}
potentials. We first discuss the results obtained by using
ultra-soft pseudopotentials. Almost all of the total magnetic
moment and the cohesive energy of the optimized structures have
been in good agreement with those given in Table~\ref{tab:3}
(obtained within collinear approximation). However, there are some
slight changes for specific finite structures. For example, Sc$_7$
is found to have magnetic moment of $7 ~\mu_B$ in collinear
approximation. Even though one of the initial condition in
noncollinear calculations resulted in the same magnetic moment and
energy, there is even a more energetic state ($0.01$ eV lower)
with total magnetic moment of $9 ~\mu_B~ (5.2,~5.2,~5.2)$. The
same situation also occurred with PAW potential. Ti$_5$ has a
special magnetic moment distribution which is the same for both
ultra-soft and PAW cases and will be explained below. In collinear
approximation, V$_5$ is noted to have zero magnetic moment,
nevertheless there is a state $0.03$ eV lower in energy which is
FM with $\mu=1~\mu_B~(0.7,~0.7,~0.2)$. Even though Co$_7$ has the
same total magnetic moment in both collinear and noncollinear
case, there is a significant energy difference between two cases.
Noncollinear structure of Co$_7$ has $\sim0.4$ eV lower energy
with magnetic moment distribution as (8.5, 8.5, 9) $\mu_B$.
Sc$_4$, Sc$_7$, Ti$_3$, Ti$_7$, V$_3$, V$_7$, Cr$_7$, all
structures of  Mn, Fe$_6$ and Ni$_5$ have truly three dimensional
(3D) magnetic moment distribution.

Finally, noncollinear (NC) calculations have been performed using
PAW potentials starting with five different initial assignment of
magnetic moments as described above. Most of our calculations have
yielded the same magnetic moment distribution with previous
calculations, but there are still few cases, which are resulted
differently. Mn$_7$ is an exception; all structure optimization
starting from different initial conditions resulted in a
non-planar geometry. Note that in collinear and noncollinear
calculations using ultra-soft pseudopotential Mn$_7$ was stable in
a local minimum corresponding to the planar zigzag geometry, but
it formed a cluster when spin-orbit coupling and NC effects are
taken into account. Unlike other $n=5$ zigzag structures, Ti$_5$
has a unique ordering of the atomic magnetic moments. Two Ti atoms
of the upper row have magnetic moments which are in opposite
direction. Similarly, two Ti atoms at the ends of the lower row
also have atomic magnetic moments in opposite direction, but the
magnitude of moments are smaller than those of the upper row. The
atom at the middle of the lower row has no magnetic moment. In
$n=6$ case, only Co$_6$ has a non-vanishing magnetic moment. Other
atoms form dimers which are coupled in the AFM order. If we assume
that the shape of $n=6$ molecule is parallelogram, there is an AFM
coupling between the atoms on both diagonals. In addition to
these, remaining two atoms in the middle also coupled in the AFM
order. Atoms in Sc$_6$, Ti$_6$ and V$_6$ have magnetic moments
only in the \emph{xy}-plane, whereas Cr$_6$, Mn$_6$ and Ni$_6$
have $3D$ magnetic moment distribution. Cr$_n$ chains exhibit an
even-odd disparity; Cr$_n$ has an AFM ordering for even $n$, but
it has a FM ordering for odd $n$. There are also cases where
collinear and noncollinear calculations with ultra-soft
pseudopotential resulted in an excited state for magnetic moment
distribution. Although PAW potential calculations found the same
magnetic ordering with collinear and ultra-soft NC cases, there
are even more energetic states for Sc$_6$, V$_4$, Cr$_5$ and
Mn$_5$ given in Table~\ref{tab:4}. Geometric dimerization also
plays an important role in determining the average cohesive energy
of the system. Interestingly the cohesive energies of V$_6$ and
V$_7$; Cr$_6$ and Cr$_7$; Ni$_4$ and Ni$_5$ are not changing with
$n$. It should be denoted that Hobbs \emph{et. al}\cite{1} carried
out noncollinear calculations with the PAW potential on Cr$_{2-5}$
and Fe$_{2-5}$ finite chain structures. Here, our results are in
agreement with those of Hobbs \emph{et. al}.\cite{1}

We calculated the spin-orbit (SO) coupling energies,
$\triangle$$E_{so}^x$ and $\triangle$$E_{so}^z$, under a unit
magnetic field along $x$- and $z$-direction, respectively. Here,
the optimized structure of every initial condition together with
the calculated magnetic moment on the individual atoms are used
for the calculation of SO coupling. The optimized structures of
(TM)$_n$ and atomic magnetic moments, $\mu_a$, have been
determined within noncollinear approximation using PAW potentials.
The results are given in Table~\ref{tab:4} in units of meV. As it
can be easily seen that SO coupling does not play an important
role on the energy of the planar finite structure. However, SO
coupling becomes crucial when the total magnetic moments, which
happen to be oriented in different directions owing to the
different initial conditions, result in the same energy. In this
case SO coupling energy difference changes with respect to the
direction of the applied magnetic field. In Table~\ref{tab:4},
only the most energetic configurations including the SO coupling
effects are given. $\triangle$$E_{so}^x$ and $\triangle$$E_{so}^z$
appear to be independent of $n$ except Mn$_5$, Co$_2$ and Co$_3$.
It is also observed that when SO coupling is taken into account,
LUMO-HOMO gap energies decreases as in the bulk structures. Only
for Ni$_4$, Cr$_5$, Mn$_5$ and Ni$_7$, LUMO-HOMO gap increased due
to the fact that the final geometry of SO coupling calculations
has further relaxed slightly from that of NC calculations. We
close this section by noting that PAW potential is found to be
more suitable for the following reasons. First, the individual
atomic character in the chain structure, as well as local magnetic
moment are better represented by the PAW potential. Secondly,
using PAW potential one can provide an accurate prediction of
spin-orbit coupling.
\\
\section{Conclusion}

In this paper, we presented an extensive study of the structural,
electronic and magnetic properties of monatomic chains of light
transition metal atoms (Sc, Ti, V, Cr, Mn, Fe, Co, Ni as well as
Cu and Zn) using first-principle plane wave methods. We considered
infinite and periodic chains (with linear, dimerized linear,
zigzag and dimerized zigzag geometry) and small chains including
2-7 atoms. Because of end effects, we found a dramatic differences
between infinite chains and finite ones. Therefore we believe that
the basic understanding of monatomic TM chains have to comprise
both infinite and finite structures as done in the present paper.

The infinite, dimerized linear structures have a shallow minimum
only for a few TM atoms; planar zigzag and dimerized zigzag
structures, however, correspond to a lower binding energy
providing stability in this geometry. As for short chains
consisting of 4-7 TM atoms, the planar zigzag structure is only a
local minimum. The finite chains tend to form clusters if they
overcome energy barriers. We found close correlation between
magnetic state and geometry of chain structure. In this study, we
presented the variation of binding energy as a function of lattice
constant for different structures and magnetic states. We also
revealed the dependence of electronic and magnetic properties on
the atomic structures of chains. We found that the geometric
structure influence strongly the electronic and magnetic
properties of the chains. For example, infinite linear Vanadium
chain becomes half-metallic upon dimerization. Similarly, infinite
dimerized and metallic Sc chain becomes half-metallic with $100\%$
spin-polarization at the Fermi level upon transformation to zigzag
structure. Furthermore, while the infinite linear Mn chain has an
AFM ground state, with $\mu=\sum \mu_a = 0$, but $|\sum
\mu_a^{\uparrow}| = |\sum \mu_a^{\downarrow}| = 4.40 ~\mu_B$, it
becomes a FM metal  with $\mu = \sum \mu_a = 4.36 ~\mu_B$ as a
result of the structural transformation from linear to dimerized
zigzag structure.

Magnetic ordering of finite-chains becomes more complex and
requires a treatment using noncollinear approximation. The
structure optimizations carried out using ultra-soft
pseudopotentials generally result in the same cohesive energy and
magnetic moment in both collinear and noncollinear approximations.
However, for specific finite chains the total magnetic moments
calculated by using PAW potentials with the same initial magnetic
moment distribution differ dramatically from ultra-soft results.
Of course, our results which covers much more than $3000$
different structure optimizations may not include the lowest
energy state, but indicates the importance of noncollinear
treatment.

\section{Acknowledgement}
Part of the computational resources for this study has been
provided through a grant (2-024-2007) by the National Center for
High Performance Computing of Turkey, Istanbul Technical
University.

\end{document}